# An Application-oriented Model for Wireless Sensor Networks integrated with Telecom Infra

*Dhruv Srivastava*

*Department of Electronics & Communication, Indian School of Mines University, Dhanbad – 826004, India, E-mail: dhruv15125@ismu.ac.in*

*Dr. Priya Ranjan*

*Assistant Professor, Department of Electrical Engineering, Indian Institute of Technology, Kanpur – 208016, India, E-mail: pranjan@gmail.com*

**Abstract**

**This paper aims to propose a significant way of remote access and real time monitoring of a particular geographic area by integrating wireless sensor clouds with existing Telecom infrastructure and applications built around them through a gateway. This utility is very potent for environment monitoring in harsh and inaccessible places like mines, nuclear reactors, etc. We demonstrate a scaled down version of multi-hop network of wireless sensor nodes and its integration with existing telecom network infrastructure via a gateway. The kind of results achieved like temperature monitoring etc. gives a glimpse of an enormous step ahead in mine safety.**

**Keywords**: Wireless Sensor Network, Telecom, Mine Safety and Gateway.

## 1. Introduction

The impact of internet, cell phones, and other communication modes has been enormous. Tremendous growth in both sensor network technology [1-3] and other applications are driving the need for new techniques and tools for analyzing and visualizing sensor data streams. Convergence of wireless sensor networks with telecommunication [4] is not far away and presents enormous possibilities for developing innovative applications for real life situations which may also bring commercial services like security updates of your home on your cell phone [5]. One of the most significant applications of wireless sensor networks is environment monitoring in harsh and inaccessible places like mines, nuclear reactors, etc [6].

Mines have a very risky and hazardous environment. Hundreds of people die every year in the mines due to lack of safety measures [7]. Workers who work in the mines are always at a risk of death. Unfortunately, the uptake of technology is slow in the mining sector. Astonishingly, things have been so terrible from the past few decades. **Table1** gives a brief glossary of recent accidents [8] that took place at various places around the world. One of the major issues in the mines especially the underground mines is the concentration of oxygen, and noxious gases like carbon monoxide, nitrogen, and methane [8,9]. The inappropriate concentration of these gases can cause suffocation and even lead to explosive conditions. If these parameters can be monitored from long distance then required actions can be taken to prevent the workers from getting into harmful situations. In general, the person responsible for managing the workers and labors who work in the mines are far away from the actual mines. These managers are more interested in keeping the mining processes going on to improve the productivity irrespective of the workers safety. By using our proposed way harmful situations in the mines can be known and the workers can be instructed accordingly avoiding danger to lives.

Integrating a wireless sensor network having gas sensors with a telecom network [4] will enable remote monitoring of the concentration of these gases and negate the level of risks to human lives. However, only temperature and light sensors were used for demonstrating the concept. Whereas, nuclear reactors require maintenance of stringent environmental conditions and thus by deploying a Wireless sensor network integrated with IP network, the conditions inside the reactor can be monitored from a remote computer.

1. Topologies for the wireless sensor network

The network topology to be selected for the wireless sensor network depends on the basis of the area to be covered, data redundancy or energy optimization issues. Like, ring topology, star topology, mesh topology etc. Every network topology has its own advantages and disadvantages. Hybrid topologies can also be implemented on the wireless sensor network. Good examples of such topologies are, star on ring topology preferred when distance coverage needs to be increased and star on star topology which is preferred to avoid data redundancy and achieve faster communication. We have implemented a tree topology on the wireless sensor network because data is collected and passed on very efficiently without much data redundancy. Also, it is suitable for enhancing the network coverage area by adding up new nodes to the network. *Figure1* shows a data flow diagram in tree topology.

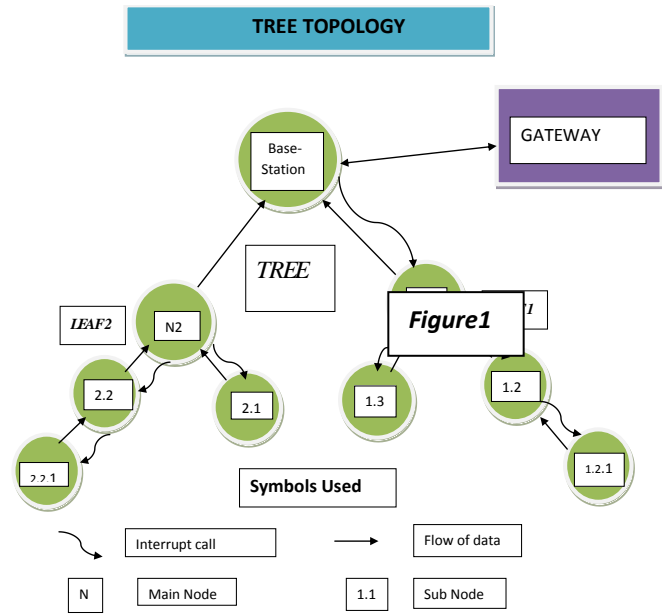

*Figure1: Data flow diagram for a wireless sensor network in tree topology*

3. Implementation of the proposed method

The wireless sensor network needs to be set up in a suitable network topology depending upon the application it has to be used for. We have implemented a tree network topology for our demonstration of our work. The network has cluster heads which collects data from every leaf node and then these cluster heads sends the whole data to the base station which is connected to an external network through a gateway. However, only temperature and light sensors were used for demonstrating the concept .Sensors, wireless technologies and computer networking [4] tools have been used. It has been demonstrated, how by our concept data from the wireless sensor network can be accessed expediently from long-distances which have been illustrated in *figure2.*

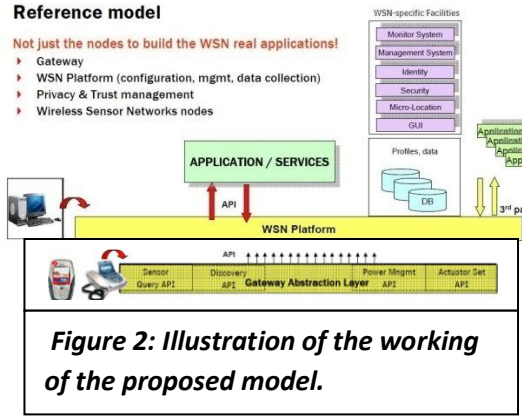

*Figure 2: Illustration of the working of the proposed model.*

As mentioned in previous section, we have used a tree topology for the wireless sensor network. The network is having 2 main leaves (cluster heads) & their 2 leaflets (sub-nodes) each. In total we monitored the values of temperature & light intensity of 6 nodes in total. **Figure 3** shows a data flow chart of what exactly have we implemented*.* We also plotted graphs for both the readings i.e. temperature & light intensity coming from every node. The graphs being plotted at the time of the experiment can be seen in *Figures 4-9.*The readings of temperature & light intensity coming from every node were taken from the base station and written onto a file (.txt). The file was being updated at regular interval of time and was sent to different clients on their request, respectively. *Table2* shows the software and hardware used for performing the experiment, hardware's detailed description has been reported [10] and can be found in their respective datasheets also.

***Terminology used:***

i. *Interrupt call-* When a node sends a message to another node asking to send the data it has.
ii. *Flow of data-*Sending information from node to another i.e. wireless communication between the nodes.
iii. *Topology-* The way/structure in which the network has been set up.
iv. *Leaf-* Used as a synonym for a node in tree topology.
v. *Leaflet-* Used as a synonym for a sub- node in tree topology.

**Assumption:** Wireless range of communication of every node is nearly same.

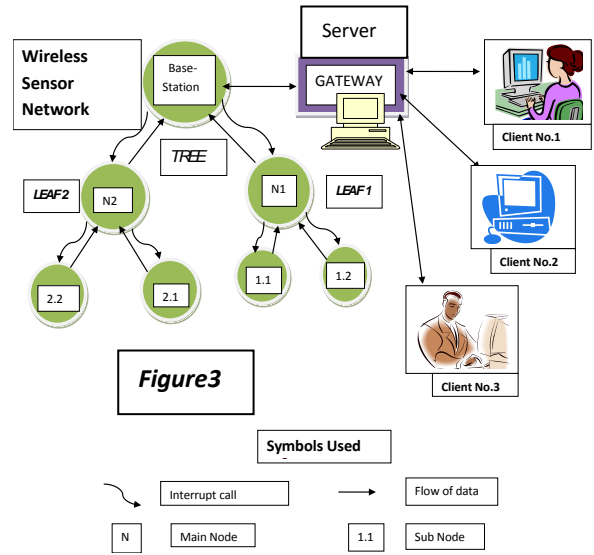

*Figure 3: The experimental set up: A flow chart of what exactly have we implemented.*

## 4. Results and Discussions

We successfully implemented a multi-hop network of sensor nodes as shown in *figure 3* which are capable of monitoring temperature, light intensity and other ambient data. The readings of temperature & light intensity coming from every node placed at various location of the laboratory were the experiment was performed. The readings were taken from the base station as it has data coming from

every node. Then the readings were plotted on graphs, which are shown in *figures 4-9.*

The base station of the deployed network is responsible for aggregation of data from all nodes and sending it to the main server via a gateway. The server can provide service to multiple clients. A customized application layer has been built on the client machines to facilitate efficient data monitoring and cluster selection from a deployed wireless sensor network. This server can provide facilities to the telecom servers and application layer interface can be built on PDAs, mobile phones, etc, thus increasing the number of clients and ease of information access. This will lead to emergence of a new generation of telecommunication value added services and provide facilities for building automation, health monitoring, advanced metering application etc.

The same setup can be used to enhance mine safety [8] as already mentioned. The wireless sensor network must be planted in a mine. The nodes of the network must be placed at various locations of the mine. Each node should be planted at different locations of the mine covering the whole area. Each node needs to be fabricated with temperature sensors like TMP-275 and various gas sensors like TGS-2611, TGS-2442, and TGS-2600 which can measure the parameters like temperature, concentration of oxygen, carbon monoxide nitrogen, and methane.

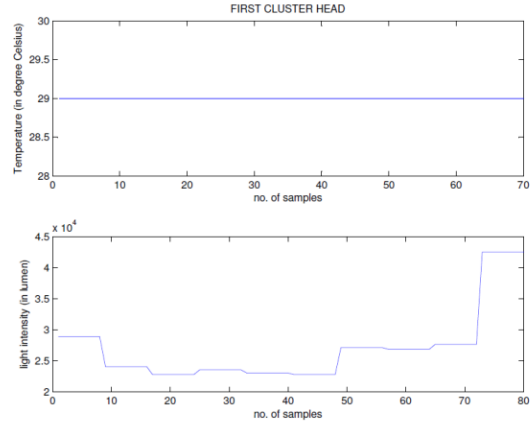

*Figure 4: Readings of node N1- first cluster head*

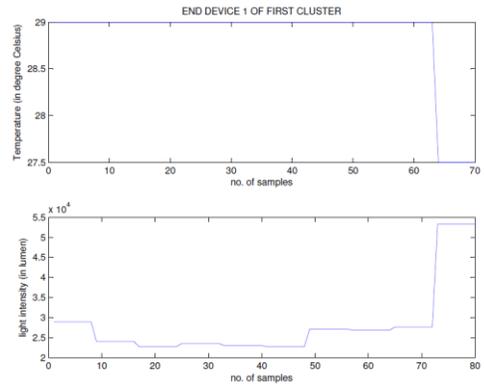

*Figure 5: Readings of node 1.1-endevice1 of first*

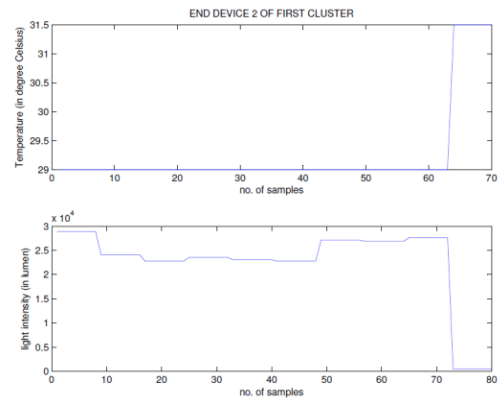

*Figure 6: Readings of node 1.2-endevice2 of first cluster head*

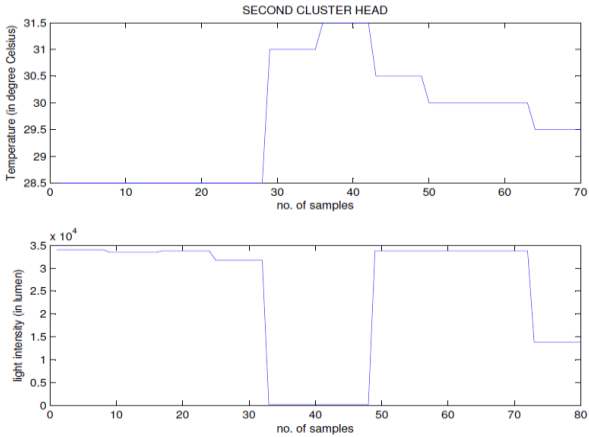

*Figure 7: Readings of node N2-second cluster head*

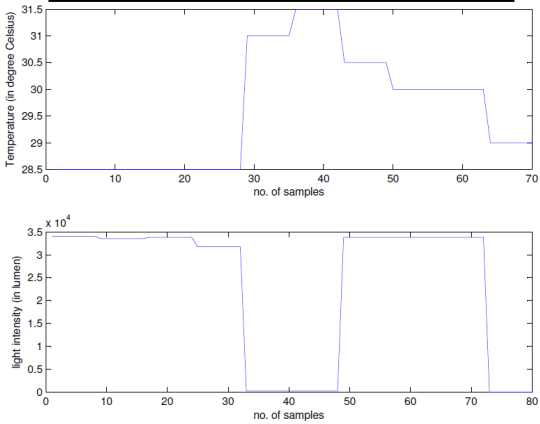

*Figure 8: Readings of node 2.1- endevice1 of second cluster head*

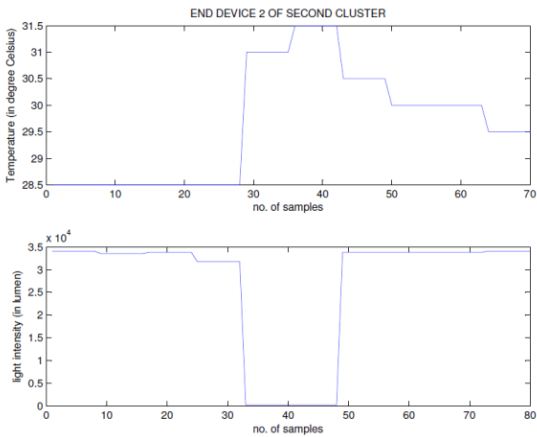

*Figure 9: Readings of node 2.2- endevice2 of second cluster head*

There is not much of variation in the temperature readings because all nodes were placed in the same room. There are a very few points where the temperature or the light intensity readings have suddenly dropped down to null. This probably happened because of the communication breakage between the nodes i.e. between the leaf and leaflets due to unknown reason. We were not able to avoid it. But important part is that the readings can be monitored and can be seen on the graphs.

## 5. Conclusions

We have successfully implemented a multi-hop network of sensor nodes which are capable of monitoring temperature & light intensity. The base station for this network is responsible for aggregation of data from all nodes and sending it to the main server via a gateway. The server then provides service to multiple clients. Though the work here has been mine safety centric the concept can open entirely new areas of business for telecom operators and web service providers, e.g. monitoring ones home while at work or on a long tour.

We already have reached a level where we have shown the effectiveness of wireless sensor network with a small set of nodes. In addition to it, most of the software made and used is open source further modifications can be made as per the requirements. We strongly believe that this work can be taken to a much bigger platform which can bring drastic changes in the present situation of mine safety and telecom sector.

## Acknowledgements

We are grateful to Mr. Vividh Mishra, Senior Project Associate, *Department of Electrical Engineering, Indian Institute of Technology, Kanpur – 208016, India*. The network programming involving server-client interaction would not have been possible without him.

## References

1. *Jeol I, Goodman, Albert I Rether and David R Martinez*, "Next generation technologies to enable sensor network" in Sensor Network protocols, Ed. Imad Mahgoub, Mohammad Ilyas, Pub. Taylor & Francis, 2006, pp2-1,2-21

2. *Xingzhen Bai, Shu Li and Juan Xu*, IETE Technical Review, V 27(2), (Apr 2010), pp 124-137

3. Sensor Handbook, Ed., *Sabrie Soloman*, Pub. McGraw Hills, 1998 Convergence of Broadcast and New Telecom Networks, Ed., *Ralf Keller, Thorsten Lohmar, Ralf Tönjes and Jörn Thielecke*

4. Convergence of Broadcast and New Telecom Networks, Ed., *Ralf Keller, Thorsten Lohmar, Ralf Tönjes and Jörn Thielecke*

5. Characteristics of wireless sensor network for full-scale ship application, Ed., *Bu-Geun Paik, Seong-Rak Cho,Beom-Jin Park, Dongkon Lee, Byung-Dueg Bae & Jong-Hwui Yun*

6. The Australian Mining and the ICT industries: productivity and growth, *A Report to NOIE and DCITA*

7. Sago Mine Accident-General Questions and Answers, Mine Safety and Health Administration (MSHA) US Department of Labor, (January 21, 2006)

8. Padhi, S.N. (2003), "Mines Safety in India-Control of Accidents and Disasters in 21st Century", *Mining in the 21st Century: Quo Vadis?,* Ed., A.K. Ghose et al., Pub.,Taylor & Francis, ISBN 90-5809-274-7

9. Annual Report , Ministry of Mines, Government of India, Pub., National Informatics Centre, (2007-2008)

10. *Dhruv Srivastava, Vaibhav Mehta , Prachi Chauthaiwale and Dr. Priya Ranjan,* Need for Mine Safety: An ICT based solution to make mines safer (Technical Report) http://arxiv.org/ftp/arxiv/papers/1007/1007.1559.pdf

*Table 1: Some recent major mining disasters*

| S. No. | Year | Country & place | Death details |
| --- | --- | --- | --- |
| 1 | * April 2010 | USA West Virginia | An explosion **killed 29** miners in in the deadliest U.S. mine disaster since 1984. |

| 2 | * February 2005 | CHINA Sunjiawan | A gas explosion at the colliery of state owned Fuxin Coal Industry Group **kills 214**. |
|---|---|---|---|
| 3 | * November 2005 | CHINA Dongfeng | A gas explosion **kills 169** people at state owned coal mine |
| 4 | * September 2006 | INDIA Jharkhand. | **50 miners are killed** after the roof of a coal mine collapses following an explosion in the eastern state of |
| 5 | * September 2006 | KAZAKHSTAN Lenin mine. | At least **41 people are killed** after an underground explosion |
| 6 | * March 2007 | RUSSIA Siberian coal mine | Blast rips through Siberian coal mine, killing at least 110 people. |
| 7 | * September 2007 | CHINA Shandong | Coal mine shaft floods in the eastern province of Shandong, killing 181 miners. |
| 8 | * September 2008 | CHINA | A mudslide caused by the collapse of a mine waste reservoir in northern China **kills 254**. |
| 9 | * November 2009 | CHINA | A gas explosion at a coal mine in northeast China kills 104. |
| 10 | * March 2010 | SIERRA LEONE | At least **200 people are killed** when a trench collapses at a gold mine in Sierra Leone |

### *Table 2: Software and Hardware used with their brief description*

| AVR Studio 4 | Software | For programming & debugging the AVR microcontroller. |
|---|---|---|
| X-CTU | Software | Allows the ability to program the radios' firmware settings via a graphical user interface.(Windows based) |
| MATLAB (optional) | Software | For serial port interfacing and plotting readings on graphs. |
| Indriya CS-03A14 Kit | Hardware | Wireless sensor module having options to plug in different combinations of sensors[1] |
| AVR Atmega 128L microcontroller | Hardware | 7.3728 MHz, 128KB flash, 4KB RAM processing unit |
| Temperature Sensor (TMP-275) | Hardware | $0.5^0C$ accuracy digital response |
| Ambient Light Sensor (APDS-9300) | Hardware | approximate human eye 16bit I2C compatible response |
| XBee Wireless Radio (2.4 Ghz) | Hardware | IEEE802.15.4 compliant radio with 30m-100m range |

| USB-UART Interface (FTDI chip) | Hardware | For communication of hardware with a computer through serial communication |